\documentclass[aps,twocolumn,prc,showpacs,preprintnumbers,
               nofootinbib,float,superscriptaddress,longbibliography]{revtex4-1}

\usepackage{epsfig}
\usepackage{color}
\usepackage[dvipsnames]{xcolor}
\usepackage{float,amsmath,amssymb,diagbox}
\usepackage{graphicx}
\usepackage{url}
\usepackage{bbold}
\usepackage{ulem}
\usepackage[utf8]{inputenc}
\usepackage{hyperref}
\usepackage{lineno}
\usepackage{tabularx}

\usepackage[utf8]{inputenc}

\begin{document}
\title{Search for baryon junctions in photonuclear processes and isobar collisions at RHIC}


\author{Nicole Lewis}
\address {Physics Department, Brookhaven National Laboratory, Upton, NY 11973, USA}

\author{Wendi Lv}
    \affiliation{State Key Laboratory of Particle Detection and Electronics, University of Science and Technology of China, Hefei 230026, Anhui, China} 

\author{Mason Alexander Ross}
\affiliation{Department of Physics, East Carolina University, Greenville, NC 27858, USA}

  \author{Chun Yuen Tsang}
    \affiliation{Department of Physics, Kent State University, Kent, OH 44242, USA}

\author{James Daniel Brandenburg}
\address {Physics Department, The Ohio State University, Columbus, Ohio 43210, USA}

\author{Zi-Wei Lin}
\affiliation{Department of Physics, East Carolina University, Greenville, NC 27858, USA}

      \author{Rongrong Ma}
    \affiliation{Physics Department, Brookhaven National Laboratory, Upton, NY 11973, USA}

\author{Zebo Tang}  
    \affiliation{State Key Laboratory of Particle Detection and Electronics, University of Science and Technology of China, Hefei 230026, Anhui, China}

\author{Prithwish Tribedy}
\email{ptribedy@bnl.gov}
\address {Physics Department, Brookhaven National Laboratory, Upton, NY 11973, USA}

\author{Zhangbu Xu}
    \affiliation{Department of Physics, Kent State University, Kent, OH 44242, USA}

    \date{\today}

\begin{abstract}
During the early development of Quantum Chromodynamics, 
it was proposed that baryon number could be carried by a non-perturbative Y-shaped topology of gluon fields, called the gluon junction, rather than by the valence quarks as in the QCD standard model. A puzzling feature of ultra-relativistic nucleus-nucleus collisions is the apparent substantial baryon excess in the mid-rapidity region that could not be adequately accounted for in most conventional models of quark and diquark transport.  The transport of baryonic gluon junctions is predicted to lead to a characteristic exponential distribution of net-baryon density with rapidity and could resolve the puzzle. In this context we point out that the rapidity density of net-baryons near mid-rapidity indeed follows an exponential distribution with a slope of $-0.61\pm0.03$ as a function of beam rapidity in the existing global data from A+A collisions at AGS, SPS and RHIC energies. 
To further test if quarks or gluon junctions carry the baryon quantum number, we propose to study the absolute magnitude of the baryon vs. charge stopping in isobar collisions at RHIC. We also argue that semi-inclusive photon-induced processes ($\gamma+p$/A) at RHIC kinematics provide an opportunity to search for the signatures of the baryon junction and to shed light onto the mechanisms of observed baryon excess in the mid-rapidity region in ultra-relativistic nucleus-nucleus collisions. Such measurements can be further validated in A+A collisions at the LHC and $e+p$/A collisions at the EIC. 
\end{abstract}

\maketitle

\section{Introduction}
The baryon number is a conserved quantum number in nature. Each quark is assigned to carry 1/3 of a baryon number in the Standard Model. The lightest baryon is the proton, which is made up of three valence quarks and carries a baryon number of one. For many decades, the standard understanding of baryons was that their valence quarks interact with gluons insides hadrons with no specific topological configuration. However, an alternative postulation~\cite{Rossi:1977cy,Artru:1974zn} in the early 1970s suggested that the valence quarks were connected in a Y-shaped structure called a gluon junction or baryon junction. This configuration is thought to be what traces the baryon number, with the junction serving as the only possible gauge-invariant structure of the baryon wave function~\cite{Kharzeev:1996sq} and having been studied in Lattice QCD~\cite{Suganuma:2004zx,Takahashi:2000te}. In most processes, the ends of the gluon junction are always connected to quarks while anti-junctions are connected to antiquarks. This makes it difficult, if not impossible, to distinguish the two scenarios. 

A puzzling feature of ultra-relativistic nucleus-nucleus collisions is the experimental observation of substantial baryon asymmetry in the central rapidity (mid-rapidity) region both at RHIC~\cite{Bearden:2003hx,STAR:2008med,STAR:2017sal} and at LHC energies~\cite{ALICE:2010hjm, ALICEabbas2013mid}. Such a phenomenon is striking, as baryon number is strictly conserved, therefore, net-baryon number cannot be created in the system and must come from the colliding target and projectile. In the picture of valence quarks carrying baryon quantum number, at sufficiently high energies one expects these valence quarks to pass through each other and end up far from mid-rapidity in the fragmentation regions~\cite{Kharzeev:1996sq,ANDERSSON198331}, seemingly inconsistent with the experimental observations. 
In the alternative picture of baryon junctions tracing the baryon number, these junctions are flux-tube configurations that contain an infinite number of gluons and typically carry a minuscule fraction of the colliding nucleons' momentum compared to the valence quarks $x_J \ll x_V$,  where $x_J$ and $x_V$ are the fraction of momentum carried by the baryon junction and valence quarks, respectively. Unlike valence quarks, the junctions from a target hadron/nucleus have sufficient time 
to interact and be stopped by the soft parton field of the projectile in the mid-rapidity region, even in high energy collisions~\cite{Kharzeev:1996sq}. While the baryon junction is stopped at a particular rapidity $y$, the valence quarks may be pulled away, producing a $q\bar{q}$ pair in the process, which will populate the region between $y$ and the fragmentation region characterized by beam rapidity $Y_{\rm beam}$. The produced baryons are expected to: 1) have low transverse momentum due to the soft partons involved in the process, 2) may have different quark content than the colliding baryons, since junctions are blind to quark flavor, and, 3) will be accompanied by many pions, therefore leading to high multiplicity events. 
However, the most important feature of the baryon-junction stopping process is the characteristic exponential drop of the cross section with the rapidity loss variable ($\sim \exp(-\alpha_J(y-Y_{\rm beam}))$) determined by the Regge intercepts of the baryon junction ($\alpha_J$) (see Ref.~\cite{Kharzeev:1996sq}). 

Conventional models, such as PYTHIA \cite{Sjostrand:2006za, Sjostrand:2007gs} and HERWIG \cite{Bahr:2008pv, Bellm:2015jjp}, which simulate $p+p$ collisions, typically employ valence quarks as carriers of baryons. When extending to simulating heavy-ion collisions~\cite{gyulassy2002proceedings,Abdel-Waged:2020mik}, various baryon production mechanisms are used, such as the ``popcorn" model in PYTHIA~\cite{ANDERSSON198331,Sjostrand:2006za}, diquarks in models like UrQMD and AMPT~\cite{bleicher1999relativistic,Lin:2004en}, or multiple strings as in HIJING~\cite{Vance:1998vh}. It is worth noting that a new baryon production mechanism through dynamical string junction formation is implemented in PYTHIA ~\cite{Bierlich:2022pfr, Christiansen:2015yqa}, which can greatly enhance baryon production at mid-rapidity, including charm baryons \cite{ALICE:2020wfu, ALICE:2021dhb}. This is realized with a color reconnection (CR) model, which  rearranges strings after initial scatterings to minimize the total string length in a collision, and string junctions emerge in the process. While junctions are involved in this new implementation, they are not present in the incoming protons, and thus do not participate in the scattering process, which is different from the mechanism proposed in~\cite{Kharzeev:1996sq}. In following sections, we will provide predictions from different heavy-ion models including HERWIG and PYTHIA models with and without CR for various observables relevant for baryon junction search.

\section{Beam Energy dependence of Inclusive net-proton yields}
Such a fundamental conjecture about baryons has never been tested conclusively in an experiment. There have been significant experimental and theoretical developments in the late 1990s and early 2000s~\cite{gyulassy2002proceedings,Kopeliovich:1998ps,arakelyan2002baryon,Takahashi:2000te}, but relatively little progress has been made in the last two decades. First, let us consider the rapidity distribution of net-baryons in $p+p$ or heavy-ion A+A collisions at a fixed energy. It is not straightforward to study the signatures of an exponentially falling cross section with rapidity as in hadronic and symmetric A+A collisions the stopping of both target ($\sim \exp(\alpha_J(y-Y_{\rm beam}))$) and projectile ($\sim \exp(-\alpha_J(y+Y_{\rm beam}))$) will contribute leading to a nearly symmetric distribution. Regardless, if one considers the production of net-baryons at mid-rapidity ($y=0$), one expects to see an exponential drop with beam rapidity, $\exp(-\alpha_J\, Y_{\rm beam})$. The exponent in Ref~\cite{Kharzeev:1996sq} was predicted to be $\alpha_J=(2-2 \alpha_0^{J})=1$ for double-baryon stopping and $\alpha_J=2-\alpha_0^{J}-\alpha_P(0)=0.42$ for single-baryon stopping using $\alpha_0^{J}\simeq 0.5$ and $\alpha_P(0)-1=0.08$~\cite{Rossi:1977cy,Donnachie:1992ny}.

\begin{figure}[htb]
  \begin{center}
    \includegraphics[width=0.5\textwidth]{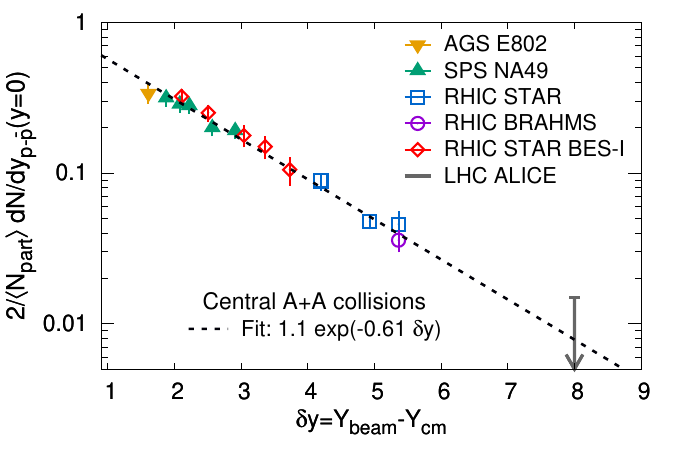}
  \end{center}
  \caption{\label{fig_baryon} Exponential dependence of mid-rapidity ($y\approx0$) net-proton density per participant pair in central heavy ion collisions with $Y_{\rm beam}$ which is equal to the rapidity difference between beam and detector mid-rapidity ($\delta y$)~\cite{STAR:2008med,STAR:2017sal,Bearden:2003hx,E802:1998cxv,NA49:1998gaz,NA49:2004mrq, NA49:2006gaj,ALICE:2013hur,ALICEPhysRevC.88.044910}}
\end{figure}

\begin{figure}[htb]
  \begin{center}
    \includegraphics[width=0.5\textwidth,trim={2cm 0.5cm 2cm 2cm},clip]{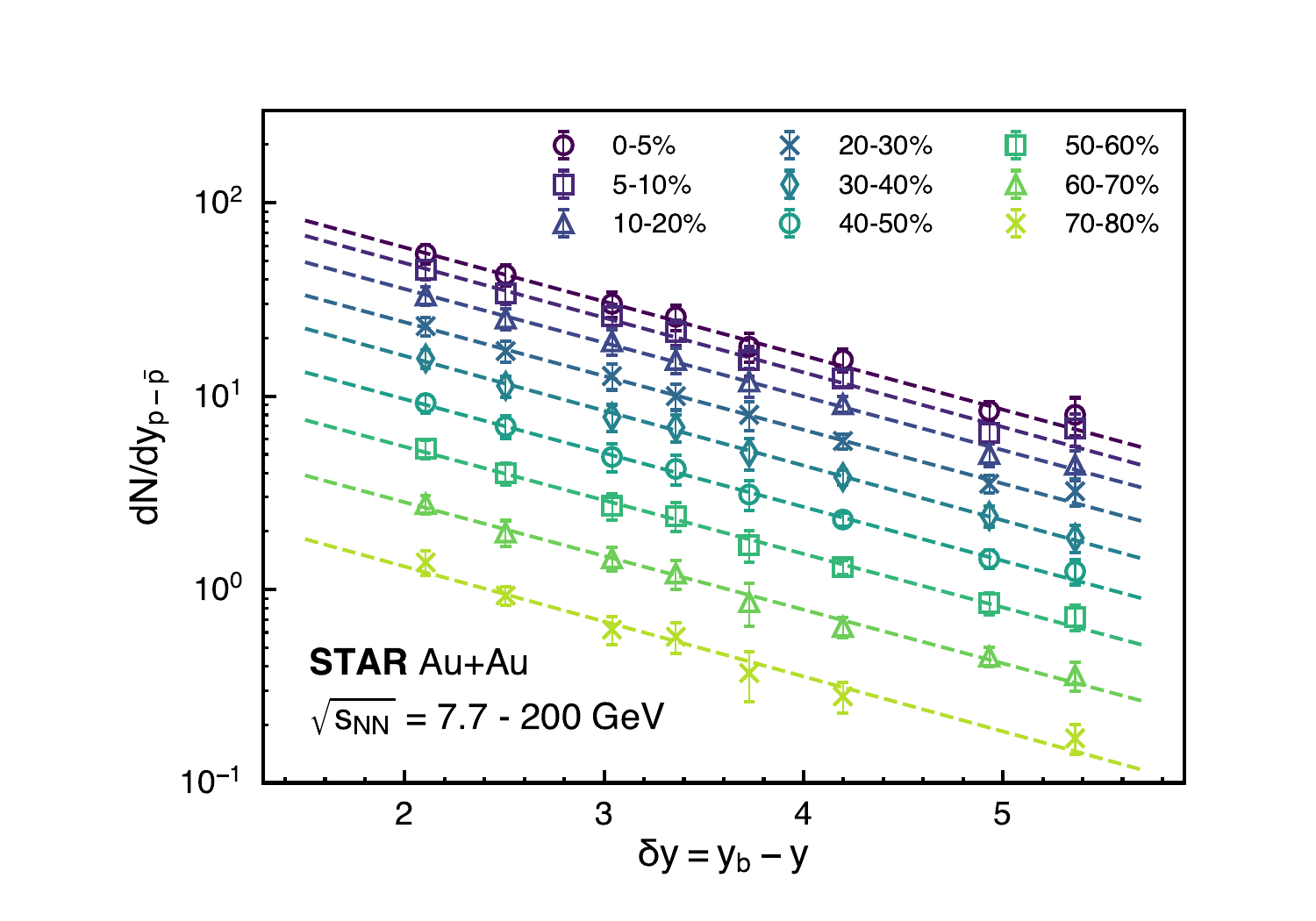}
  \end{center}
  \caption{\label{fig_baryon_allcent} Exponential dependence of mid-rapidity ($y\approx0$) net-proton density per participant pair in mid-central heavy ion collisions with $Y_{\rm beam}$ which is equal to the rapidity difference between beam and detector mid-rapidity ($\delta y$)~\cite{STAR:2008med,STAR:2017sal,ALICE:2013hur,ALICEPhysRevC.88.044910}}
\end{figure}

Global data on inclusive net-proton yields at mid-rapidity from the Alternating Gradient Synchrotron (AGS)~\cite{E802:1998cxv}, the Super Proton Synchrotron (SPS)~\cite{NA49:1998gaz, NA49:2004mrq, NA49:2006gaj},  RHIC~\cite{STAR:2008med,STAR:2017sal,Bearden:2003hx} and LHC~\cite{ALICE:2013hur,ALICEPhysRevC.88.044910} are presented in Fig.~\ref{fig_baryon} and they indicate that for central heavy-ion collisions the mid-rapidity net-baryon density follows an exponential distribution with the variable $\delta y=Y_{\rm beam}-Y_{\rm cm}$, where $Y_{\rm beam}$ is the beam rapidity and $Y_{\rm cm}$ is the center-of-mass rapidity. This variable $\delta y$ can be referred to as the ``rapidity loss" which for mid-rapidity protons produced in a collider experiment is equal to beam rapidity: $\delta y=Y_{\rm beam}$ as $Y_{\rm cm}=0$. 
A single collision energy therefore gives rise to a single point on Fig.~\ref{fig_baryon}. The published SPS NA49 data have been corrected for weak decays, where the contribution of weak decay protons was estimated to be about 20-25\%~\cite{NA49:2004mrq}. To allow these results to be compared to inclusive data, they have been multiplied by a factor of 1.25. The LHC ALICE arrow represents a 90\% confidence limit on the net proton yield estimated using the central proton yield and $\bar{p}/p$ ratio, $dN_{p-\bar{p}}/dy \approx (1- \bar{p}/p)dN_p/dy$, which was then multiplied by 1.35 since these yields have been corrected for weak decays which contribute as much as 35\%~\cite{ALICEPhysRevC.88.044910}.  All of the other included measurements use inclusive proton yields which include weak decay products.   The displayed uncertainties are those from the statistical and systematic uncertainties combined in quadrature.  The dotted line is an exponential fit to the data: 
\begin{equation}
\frac{ dN_{p-\bar{p}/}/dy }{N_{part}/2} = N_B \exp(-\alpha_B\, \delta y)
\end{equation} 
which yields 
$N_B=1.1\pm0.1$ and $\alpha_B=0.61\pm0.03$. The same net-baryon densities at mid-rapidity for various centralities using the available data from the RHIC BES program~\cite{STAR:2008med,STAR:2017sal} are shown in Fig.\ref{fig_baryon_allcent}. 
The qualitative behavior of the distributions of net-baryons with rapidity loss $\delta y$ does not change with centrality. Also, the exponent $\alpha_B$ obtained from the fit is found to be consistent across various centralities: $-0.63 \pm 0.052$ (10-20\%), $-0.67 \pm 0.059$ (30-40\%), and $-0.65 \pm 0.082$ (50-60\%).
A number of outstanding questions arise by looking at Fig.~\ref{fig_baryon} and Fig.\ref{fig_baryon_allcent}. %
What is the underlying process that led to non-zero net-protons at mid-rapidity?  
Why do we see an exponential drop of net-proton density as we move away from beam rapidity? 
What are the implications of the exponent $\alpha_B$? 

We argue that the trend of global net-proton density data from A+A collisions shown in Fig.~\ref{fig_baryon} is consistent with the baryon-junction picture and that the extracted exponent of $\alpha_B = 0.61\pm0.03$ is qualitatively consistent with the value of $\alpha_J\simeq0.42-1$  predicted by Regge theory. 
The small difference could arise from several other effects related to multiple hadronic interactions in central A+A collisions. 

A recent modeling of heavy-ion collisions indicates that indeed the inclusion of the aforementioned baryon junction is essential for describing mid-rapidity net-proton density at RHIC~\cite{Shen:2022oyg}. Clearly some of the earlier implementations of baryon junctions in the HIJING/B~\cite{Vance:1998vh,ALICE:2010hjm} (HIJING/B$\bar{B}$~\cite{ToporPop:2007df}) and other estimates~\cite{Kharzeev:1996sq}, which attempted to match the earlier experimental data with certain parameter tunes ($\alpha_J\simeq0.5)$, do not reproduce the experimental results presented in our Fig.~\ref{fig_baryon} and other measurements~\cite{ALICE:2010hjm}. This gives us a necessary impetus to investigate this further and to perform a series of more conclusive tests of the baryon junction conjecture in an experimental and data-driven way. 

The baryon stopping is often characterized by the average rapidity loss~\cite{busza1984nuclear,BRAHMS:2009wlg}, which shows the complicated beam energy dependence and is usually skewed by the large proton yields close to beam rapidity. It was concluded~\cite{BRAHMS:2009wlg} that the ``rapidity loss" of projectile baryons at RHIC breaks the linear scaling observed at lower energies. Another way of characterizing the baryon stopping is to use the $\bar{p}/p$ ratio~\cite{STAR:2001rbj,ALICE:2010hjm, ALICEabbas2013mid}. Both pair production and baryon stopping contribute to this ratio. The pair production grows exponentially ($\alpha_P-1$)\cite{Kharzeev:1996sq} with $\delta y$ while baryon stopping decreases exponentially ($\alpha_B$). One would expect $1/R=p/\bar{p}=1+C_1\exp{(-(\alpha_B+(\alpha_P-1))\delta y)}=1+C_1\exp{(-0.69\delta y)}$ using our fit result of $\alpha_B=0.61$ and $\alpha_P-1=0.08$\cite{Rossi:1977cy,Donnachie:1992ny}. The ALICE Collaboration~\cite{ALICEabbas2013mid} introduced a form of $1/R=p/\bar{p}=1+C_1\exp{((\alpha_J-\alpha_P)\delta y)}=1+C_1\exp{(-0.7\delta y)}$ to study baryon stopping in p+p collisions. Although the equations are quite different, accidentally the numerical values are surprisingly close to each other. It was argued that ``the results are consistent with the conventional model of baryon-number transport"~\cite{ALICE:2010hjm} and ``these dependencies can be described by exchanges with the Regge-trajectory intercept of $\alpha_J=0.5$"~\cite{ALICEabbas2013mid}. The ALICE results also disfavor any significant contribution of an exchange not suppressed at large $\delta y$ (reached at LHC energies). However, none of these aforementioned findings at the LHC is inconsistent with the observation of Fig.\ref{fig_baryon} and rules out the baryon junction picture. 

The LHC experiments have not yet obtained non-zero net-proton yield in A+A collisions.  The central values are much smaller than the uncertainty with both positive and negative net-proton yields in different centrality bins~\cite{ALICEPhysRevC.88.044910}. An upper limit of 90\% Confidence Level derived from their proton and antiproton yields is consistent with the extrapolation shown in Fig.~\ref{fig_baryon}. This extrapolation could be used to estimate the $\bar{p}/p$ ratios at various energies and we obtained $\bar{p}/p\simeq$0.95 at 0.9 TeV and 0.99 at 7 TeV. These values are quite compatible with the measurements in p+p collisions at the LHC~\cite{ALICE:2010hjm}. At this point, we would conclude that the available measurements at the LHC are not inconsistent with the baryon junction picture. However, as argued in Ref~\cite{Kharzeev:1996sq}, the gluon junction is a compact object and may be much more relevant in central A+A collisions than the $p+p$ collisions. Future LHC heavy-ion runs with improved PID and reduced systematic uncertainties ($<<\pm1\%$) could be an important venue for verifying the exponential extrapolation of the trajectory. 

Putting all the numbers together one expects the cross section of single baryon stopping to follow a similar exponential dependence of rapidity in $p+p$ collisions~\cite{Kharzeev:1996sq}. 
Clearly, this exponential dependence predicted by Regge theory is a verifiable signature of the stopping of baryon junctions. 
We argue that photon induced interactions on hadrons and nuclei provide an opportunity in this context. First, the photon is the simplest object which may fluctuate into a single dipole to interact, to first order, with only a gluon, a quark, or a baryon junction. Secondly, due to the absence of baryons in one of the colliding objects the characteristic exponential shape may be visible in $\gamma+p$ and $\gamma+A$ interactions -- something that can be tested in ultraperipheral collisions (UPCs) at RHIC and at the EIC. Indeed, if the baryons are carried by the gluon junctions and not by valence quarks, there would be a measurably smaller amount of charge stopping than baryon stopping. 
We propose to measure this effect using the RHIC isobar collisions which changed the colliding nucleus charge from Ruthenium (Ru) with ($Z=44$) to Zirconium (Zr) with ($Z=40$) without changing the number of baryons ($A=96$)~\cite{STARCME:2021mii}. 
The key feature of the isobar collisions is that the detector acceptance and efficiency all cancel out between these two colliding systems and therefore allow us to detect very small differences in the charge stopping by changing the charge of the initial nuclei~\cite{XZB2022}. Note that in this study we refer to electric charge as ``charge" for simplicity.

\section{Inclusive photon-induced processes at RHIC}

 \begin{figure*}[htb]
  \begin{center}
    \includegraphics[width=0.9\textwidth]{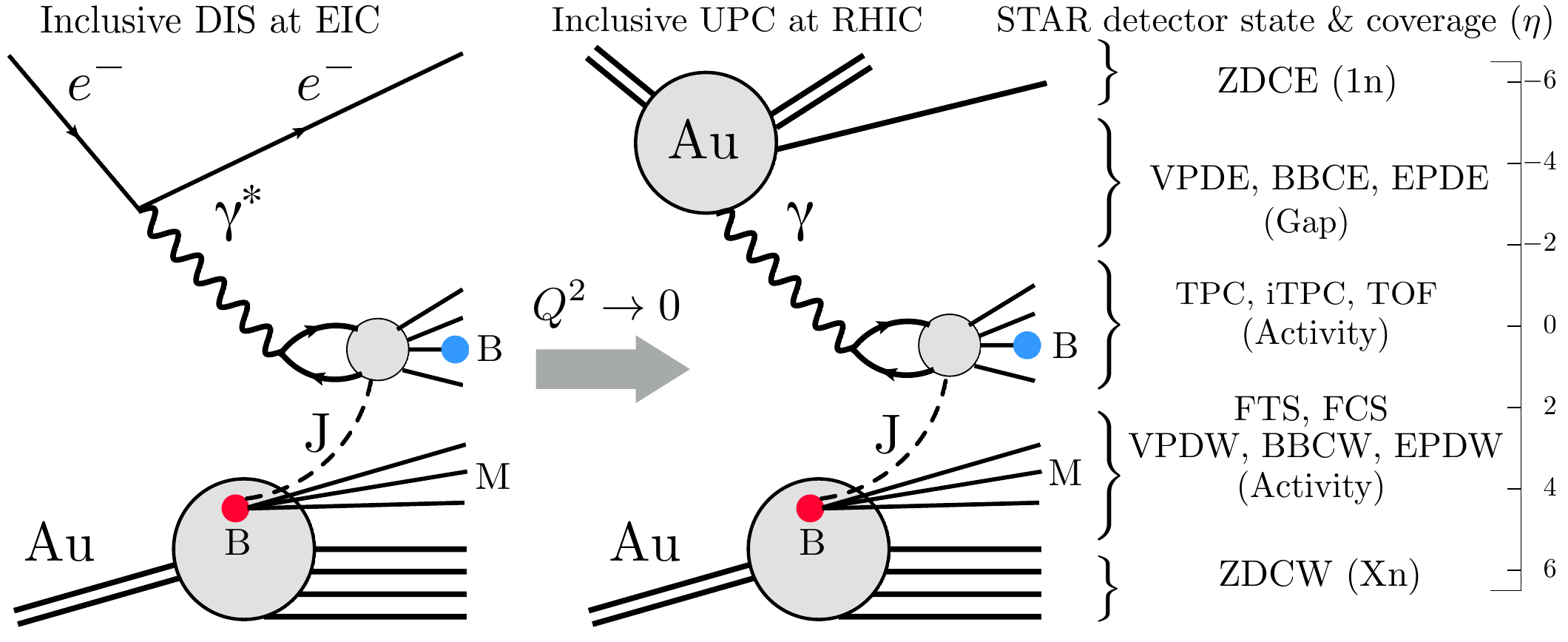}
  \end{center}
  \caption{\label{fig_rhic2eic} (Left) A cartoon of inclusive deep inelastic scattering processes ($\gamma^{*}+p$/A$\rightarrow$ X) at HERA or at the future EIC in $e+p$/Au collisions. (Middle) The low virtuality limit of deep inelastic scattering (photoproduction) that can be studied by triggering ultra-peripheral heavy ion collisions ($\gamma+p$/d/Au$\rightarrow$ X) at RHIC. Both left and middle panels depict the scenario when an incoming baryon (B) from the target ion can be stopped by the incoming photon near the central rapidity by exchanging the baryon junction (J) while the original quarks fragment as mesons (M) filling up the gap between mid-rapidity and beam fragmentation. The flavor of the baryon at mid-rapidity (shown in blue) can be different from the one in the incoming target (shown in red) as junctions are flavor-blind. (Right) The acceptance and status of different detector systems in STAR that will be active or see gap in these processes. In this case of inclusive UPCs, the photon-emitting ion may get Coulomb excited to emit a single neutron ($1n$) that will be detected by one side of the ZDCs, while the target ion will fragment into many neutrons ($Xn$) that will fill the other ZDC.}
\end{figure*}

\begin{figure*}[htb]
  \begin{center}
    \includegraphics[width=1.0\textwidth]{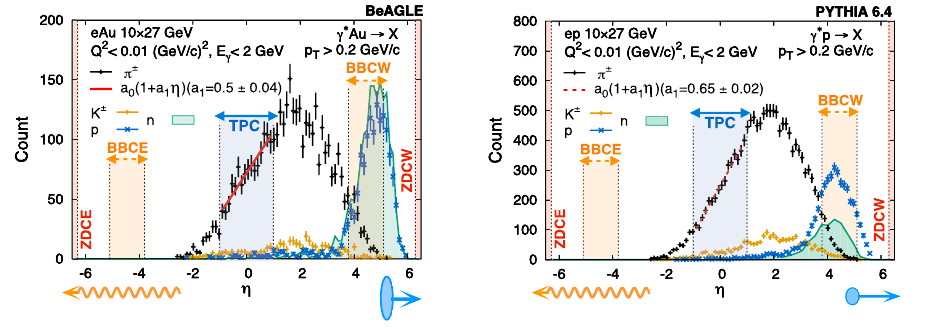}
  \end{center}
    \caption{(Left) Pseudorapidity distribution of particles in $\gamma$+Au interactions simulated using the BeAGLE event generator~\protect\cite{Chang:2022hkt, beagle, Tu:2020ymk} in the low virtuality limits ($Q^{2}<0.01 ~\rm{GeV}/c^2$) of $e+$Au DIS by restricting the energy of photons to be $E_\gamma<2 ~\rm{GeV}$ and with ion energy of $27 ~\rm{GeV}$. The pseudorapidity distributions thus produced are used to apply cuts on detectors in STAR to identify $\gamma$+Au candidates in Au+Au UPCs at $\sqrt{s_{_{NN}}}=54 ~\rm{GeV}$. (Right) The same simulated using the PYTHIA event generator for $e+p$ DIS showing qualitatively similar trend. For these processes the photon-going direction is chosen to be negative rapidity. The vertical color bands show the acceptance of different detector systems available during the collection of Au+Au UPC dataset at $\sqrt{s_{_{NN}}}=54 ~\rm{GeV}$.}\label{fig_dndeta_photonuclear}
\end{figure*}

Fig.~\ref{fig_rhic2eic}(left) shows the diagram of inclusive deep inelastic scattering (DIS, $\gamma^{*}$+$p$/A$\rightarrow X$) at HERA and at the future EIC in $e$+$p$/A collisions. Processes with virtuality of the exchanged photons $Q^2>1 ~\rm ({GeV/c})^2$ are referred to as DIS, but the majority of $e+p/A$ collisions have $Q^2$ much less than $1 ~\rm ({GeV}/c)^2$ and are instead referred to as photoproduction processes~\cite{ZEUS:2017nkv}. Such photoproduction processes in $\gamma^{*}$+Au can also be studied in UPCs at RHIC and LHC. Fig.~\ref{fig_rhic2eic} (right) shows the typical kinematics for UPCs at RHIC. For the STAR experiment, UPC datasets with photonuclear processes are available for Au+Au collisions at center of mass energy per nucleon-nucleon pair $\sqrt{s_{_{NN}}}=54$ and $200 ~\rm{GeV}$. 
In UPCs the gold ions are the source of quasi-real photons. 
The size ($R_{A}\sim 1.2 \, A^{1/3}$ fm) and charge ($Z=79$) of gold ions (mass number $A=197$) and the Lorentz boost $\gamma_{L}=27-100$ at RHIC determines the energy of the quasi-real photons $E_\gamma=\gamma_{L}(\hbar c/R_A)=0.8-2.8 ~\rm{GeV}$. The virtuality and transverse momentum are $Q^2{}^<_{\sim} \, (E_{\gamma}/\gamma_{L})^2 \simeq(\hbar c/R_A)^2=0.0008 ~\rm{GeV}^2$. The typical range of the center of mass energy of the photon-nucleon system is $W_{\gamma N}=\sqrt{4 E_\gamma E_A}\approx9-34 ~\rm{GeV}$ for $\sqrt{s_{_{NN}}}=2 \, E_{A}=54-200~\rm{GeV}$. These numbers are close to what are quoted in Ref~\cite{Baltz:2007kq}. However, it is not straightforward to estimate the range of the momentum fraction of the partons on which the photon scatters (Bjorken-x) in these interactions as they are process dependent. Due to limited control, the distribution of various kinematic parameters in UPCs, particularly $x$ and $W_{\gamma N}$, can only be estimated using Monte-Carlo models or more sophisticated data-driven methods~\cite{ATLAS:2021jhn}.

In the majority of cases the quasi-real photon fluctuates into a $q\bar{q}$ system (shown by the particles coming from the photon ($\gamma$) in Fig.~\ref{fig_rhic2eic}) that scatters with the partons in the target nucleus, this is referred to as a resolved process~\cite{Bertulani:2005ru}. If the baryon-junction picture is valid the following can happen: 
The $q\bar{q}$ system can interact with the baryon junction (J) inside an incoming baryon (proton or neutron, shown by a red dot in Fig.~\ref{fig_rhic2eic}) of the target ion. Such an interaction may slow down or excite the junction at mid-rapidity. This junction will eventually acquire new quarks from the vacuum and become a baryon of different flavor (shown by a blue dot in Fig.~\ref{fig_rhic2eic}) as junctions are flavor blind. 
This process will lead to production of additional mesons. 
Note that there will be enough time available for both the $q\bar{q}$ system and the junction to interact with each other since both of them carry a much smaller fraction of longitudinal momentum compared to the valence quarks. 
As a result, the original valence quarks of the incoming baryon will fragment as mesons filling up the gap between the target and mid-rapidity. 
The details of the interaction between the $q\bar{q}$ system and the junction (depicted as a blob in Fig.~\ref{fig_rhic2eic}) will determine the cross-section of this process. However, it is expected that since the projectile ($\gamma$) is baryon-free, the stopping of baryon in $\gamma+$A processes will lead to a clear asymmetric dependence of net-baryon production with $y-Y_{\rm beam}$ that can be tested in experiment~\cite{Artru:1990wq}. In hadronic collisions, Regge theory predicts a symmetric dependence of net-baryon to be $\exp(\alpha_J(y-Y_{\rm beam}))+\exp(-\alpha_J(y+Y_{\rm beam}))$ with $\alpha_J\approx 0.5$. In the most simplistic picture one expects to see an asymmetric dependence of $\exp(\alpha_J(y-Y_{\rm beam}))$ in $\gamma+A$ processes, where  $\alpha_J$ can be measured and directly compared to predictions from Regge theory. Therefore, the lower the target energy is, the more measurable the net-baryon yield is expected to be at mid-rapidity. 

Although we have limited control over the kinematics as compared to $e+p$/Au DIS or photoproduction, UPCs provide the best shot for studying inclusive quasi-real photon-induced processes ($\gamma$+A$\rightarrow$ X) off nuclei before the EIC era. Although UPCs have been studied for a long time, measurements of multi-particle production by triggering on high activity inclusive photonuclear processes have only started recently. Such an effort requires a high statistics data sample as well as large acceptance detectors with tracking and particle identification capabilities. The search for collectivity in photonuclear processes has been already initiated at the Large Hadron Collider by the ATLAS and CMS collaboration~\cite{ATLAS:2021jhn,CMS:2022doq} and modeled using  UrQMD in Ref~\cite{Zhao:2022ayk}. However, an  experimental test of baryon conjecture remains untested due to limitations of particle identification capabilities. Due to higher collision energies, the target beam rapidity is large for experiments at the LHC. This leads to a smaller measurable baryon asymmetry at the central rapidity and constitutes a major challenge for these measurements. Therefore, the RHIC UPC program provides a unique opportunity in this context.

\section{Triggering inclusive photon-induced events}

Monte-Carlo simulations using PYTHIA 6.4 ($e+p$), BeAGLE ($e$+Au), and UrQMD (Au+Au) event generators indicate several challenges in identifying inclusive $\gamma+p$/Au interactions, which remains largely unexplored at RHIC~(see Ref.~\cite{cfns2021}). In this section we discuss how such processes can be identified using the STAR detector.  Fig.~\ref{fig_dndeta_photonuclear} shows the pseudorapidity ($\eta$) distribution of identified particles with transverse momentum $p_{T}>0.2 ~\rm{GeV}/c$ in inclusive $e$+Au DIS ($\gamma^*$+Au, $\gamma^*$ refers to a virtual photon) processes simulated using the EIC Monte Carlo BeAGLE~\protect\cite{Chang:2022hkt,beagle,Tu:2020ymk} with electron and ion beam energies of $10$ and $27 ~\rm{GeV}$, respectively. The same is also repeated for inclusive $e$+p DIS using a PYTHIA 6.4 simulation~\cite{Sjostrand:2006za} resulting in very similar trends in the distributions. In these simulations, the virtuality of the exchanged photon is restricted to be $Q^2<0.01 ~\rm ({GeV}/c)^2$ and photon energy is restricted to be $E_\gamma<2 ~\rm{GeV} $ to mimic $\gamma$+Au interactions in Au+Au UPCs at $\sqrt{s_{_{NN}}}=54$ and $200 ~\rm{GeV}$. From our PYTHIA analysis we find that for our kinematic cuts about $46\%$ of the $\gamma^*+p$ process are leading order DIS ($\gamma^*+q\rightarrow q$). About $3\%$ processes are resolved process in which the photon fluctuates into an object such as a $q\bar{q}$ pair. A large fraction of event ($48\%$) are mediated by processes where the photon fluctuates into a vector meson.


Fig.~\ref{fig_dndeta_photonuclear} shows the case in which the photon emitting ion was going in the negative rapidity, or east-going, direction. On the same figure we also show some subsystems of the STAR detector, such as Time Projection Chamber (TPC) covering  $-1.0<\eta<1.0$~\cite{Anderson2003659}, the Zero Degree Calorimeters (ZDC) covering $|\eta|>6.3$ ~\cite{Adler:2000bd}, and Beam-Beam Counters (BBCs) covering $3.8<|\eta|<5.1$~\cite{Whitten:2008zz}. We argue that given such combination of detectors at mid and forward rapidities one can trigger and analyze inclusive photon-induced processes. As shown in Fig.~\ref{fig_dndeta_photonuclear} (left) a large amount of activity is seen in the west side BBC due to fragmentation protons while a gap is seen in the east side BBC. The pseudorapidity distribution of charged tracks that are mostly pions in the TPC is strongly asymmetric -- this is something that can be measured and compared to model predictions. Most importantly the west side ZDC will see a few neutrons from the fragmenting ion while the east side ZDC will only see one or two neutrons due to Coulomb excitations that is not incorporated in these simulations. In terms of triggering the $\gamma$+Au interactions the most stringent selection criterion is that the ZDC east (ZDCE) detector should be restricted to have a single neutron hit (1n), while the ZDC west (ZDCW) is required to have more than one neutron (Xn) to trigger on $\gamma$+Au candidates with east-going photons, and vice versa. This additional requirement of single neutron hit (1n) instead of requiring zero neutrons (0n) in the photon-going direction can help us to reduce backgrounds from fixed target events and beam-gas events which also produce pseudorapidity asymmetry and can mimic $\gamma$+Au interactions. With such a cut applied, additional asymmetric cuts on the BBCs help to improve the purity of our $\gamma$+Au interactions. For an ion beam energy of $27 ~\rm{GeV}$ the beam rapidity will be $Y_{\rm beam}=4.06$ will fall inside the BBC acceptance $3.8<|\eta|<5.1$. The BBCs, therefore, measure the beam fragments, spectators and can become extensions of the ZDCs for this kinematics. 



\section{Test of baryon junction hypothesis with photonuclear interactions}
\begin{figure*}[htb]
  \begin{center}
    \includegraphics[width=1.0\textwidth]{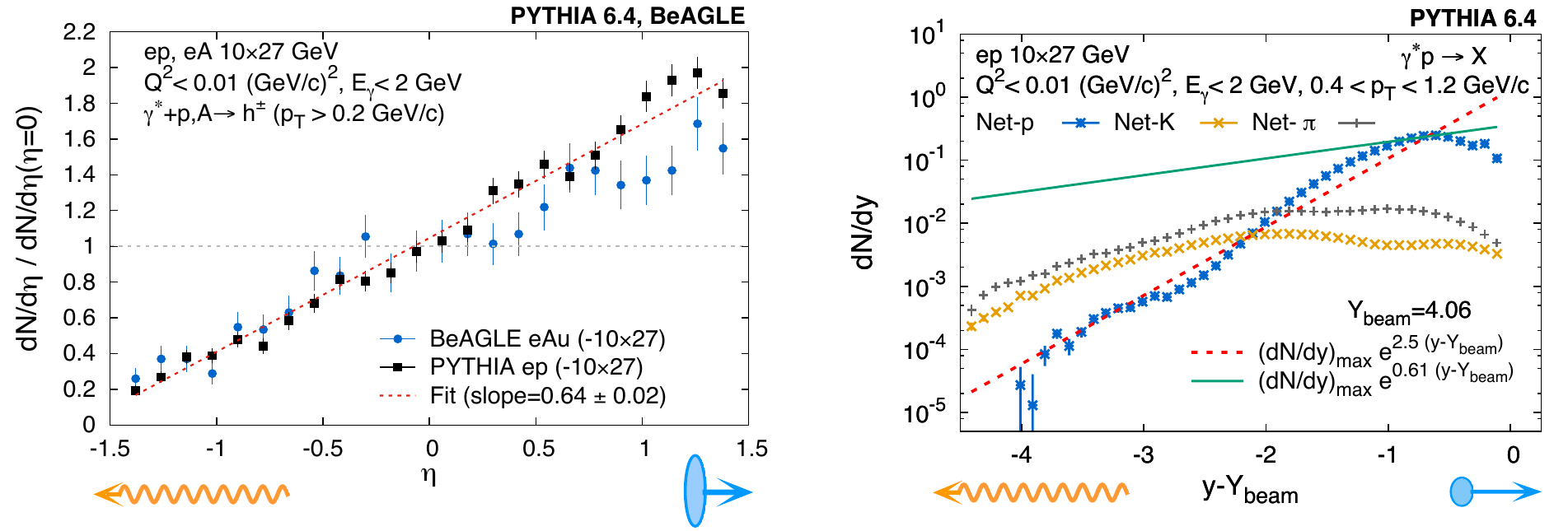}
  \end{center}
  \caption{\label{fig_baryon_pythia}(Left) Pseudorapidity distribution of inclusive charged hadrons normalized by the same at zero pseudorapidity. Results are shown for $\gamma+$Au events simulated by BeAGLE and PYTHIA for $e+$Au and $e+p$ in the limits of small photon virtuality ($Q^2\rightarrow$0) and energy ($E_{\gamma}<2$ GeV); they have clearly asymmetric distributions. (Right) Rapidity distributions of net-particles from the PYTHIA model that does not include baryon junctions. The dashed red and solid green lines are shown to guide the eyes on the $y-Y_{\rm beam}$ dependence of the distributions. A strong dependence for net-protons is predicted by PYTHIA compared to what is expected for the stopping of baryon junctions. For both panels the photon and proton/ion going directions are along negative and positive rapidities, respectively.}
\end{figure*}

The first indication of a successful selection of $\gamma+$Au processes will be an asymmetric pseodorapidity distribution ($dN/d\eta$) of inclusive charged hadrons at mid-rapidity. This is shown in Fig.~\ref{fig_baryon_pythia}(left) from simulations of inclusive $e+p$, $e$+Au collisions mimicking UPC kinematics using PYTHIA 6.4 and BeAGLE event generators, respectively. Both the simulations predict similar distributions of $dN/d\eta$ when normalized at $\eta=0$. In contrast heavy ion collision events are expected to produce symmetric distributions $dN/d\eta$ with $\eta$.

Once an enriched sample of $\gamma+$Au processes are collected in experiment, it is important to discuss the expectations of net-baryon distributions in such processes. For this we will consider net-protons as a proxy for net-baryons and use a PYTHIA 6.4 simulation that does not include the implementation of baryon junctions. In $\gamma+$Au processes, since the photon fluctuates into a dipole of quark and antiquark, it would interact with one quark or a gluon junction creating mesons or baryons at mid-rapidity at the first-order ($2\rightarrow 2$) approximation. In this picture one would expect very little baryon stopping relative to the Au+Au collisions in the absence of a baryon-junction mechanism as shown in the PYTHIA simulation in Fig.~\ref{fig_baryon_pythia}(right). In this figure the  distributions of net particles (net-pion, net-kaon, and net-protons) are plotted with $y-Y_{\rm beam}$ for $\gamma+p$ collisions. Here the photon-going direction is chosen to be along the negative rapidity direction. A much steeper distribution is observed for net-protons as compared to net-pion and net-kaon distributions. To guide the eye, two solid lines are drawn to demonstrate that PYTHIA 6.4 predicts a rapdity dependence of net-proton which is close to $\exp(2.5(y-Y_{\rm beam}))$ as compared to $\exp(0.61(y-Y_{\rm beam}))$ that is expected based on the global A+A data shown on Fig.\ref{fig_baryon}. The PYTHIA 6.4 distribution deviates from a perfect exponential, displaying peak-like structures within the range of $-4.06 < y-Y_{\text{beam}} < -2.86$ and around $y-Y_{\text{beam}} \approx -0.66$. Consequently, fitting the distribution with a single exponential function across the entire $y-Y_{\text{beam}}$ range shown in Fig. \ref{fig_baryon_pythia} is challenging. To provide a more quantitative estimate of the net-proton rapidity distribution's slope, we attempted to fit the PYTHIA 6.4 results using the function $A \times \exp(-\alpha_B (y-Y_{\text{beam}}))$, adjusting the fit range accordingly. The fit within $-4.06 < y-Y_{\text{beam}} < -2.86$ yields a slope parameter $\alpha_B = 2.18 \pm 0.12$ with a $\chi^2/\text{NDF} = 2.39$. Conversely, fitting the range $-2.86 < y-Y_{\text{beam}} < -1.76$ results in a slope parameter $\alpha_B = 3.81 \pm 0.01$, accompanied by a high $\chi^2/\text{NDF} = 9.66$, suggesting that an exponential function does not adequately describe this range.

In addition, we employ the advanced PYTHIA 8.3 version with color reconnections (CR Mode 2) beyond the leading-color approximation as outlined in Refs.~\cite{Bierlich:2022pfr, Christiansen:2015yqa}. Valence quarks remain the primary carriers of baryons, but PYTHIA 8.3 further simulates baryon formation during fragmentation by generating string junctions, enhancing baryon transport near mid-rapidity. We simulate $\gamma+p$ collisions to analyze the slope of the net-proton distribution. Our findings show a less steep distribution than PYTHIA 6.4. Fitting the PYTHIA 8.3 distribution within the range of $-5.16 < y-Y_{\text{beam}} < -1.96$ gives a slope parameter $\alpha_B = 2.71 \pm 0.02$ with a $\chi^2/\text{NDF} = 1.86$. Modifying the upper limit of fitting from -1.96 to -2.96 causes a variation in $\alpha_B$ between $2.46$ and $2.76$ (and $0.95 < \chi^2/\text{NDF} < 1.92$).

The experimental measurements of baryon stopping by selecting $\gamma+$Au processes at RHIC are a work in progress. Recently, the STAR Collaboration~\cite{Lewis:2023nmd} presented the preliminary results on $\bar{p}/p$ ratio in $\gamma+$Au relative to peripheral Au+Au collisions at small transverse momentum at the Quark Matter 2022 conference. It indicates more baryon stopping in $\gamma+$Au collisions. Further studies are ongoing in this direction. 
Studying the dependence of the baryon rapidity shift as a function of rapidity relative to the ion beam rapidity and, if possible, at different beam energies would uniquely identify the stopping mechanism, and therefore provide insights into the carrier of the baryon number.

\section{Test of baryon junction hypothesis with isobar collisions} 

The valence quarks carry electric charge, and the question is whether at the same time they also carry the baryon quantum number. One of the most straightforward investigations of whether valence quarks carry baryon number is to study the correlations of charge and baryon stopping in A+A collisions. Recent measurements of the Breit-Wheeler process in A+A collisions at RHIC~\cite{STAR:2018ldd} ($\gamma+\gamma\rightarrow e^+e^-$) and the LHC~\cite{ATLAS:2018pfw} ($\gamma+\gamma\rightarrow \mu^+\mu^-$) show that the experimental measurements, even in violent A+A collisions, match well with the QED calculations~\cite{Brandenburg:2021lnj}. Such QED calculations are performed with the assumption that projectile and target nuclear charge distributions maintain their trajectory and velocity throughout the course of the collisions. This seems to point to the possibility of a small charge stopping at the initial stage. 

It is actually challenging to perform an experimental measurement of charge stopping. It was proposed in the 1990s to use the forward bremsstrahlung to measure the charge stopping at the initial impact~\cite{Jeon:1998tq}. And while this was a creative idea, it is a very difficult proposal for experiments without a successful follow-up. For a recent proposal in this direction, we refer the readers to Ref~\cite{Park:2021ljg}. Another possibility is to directly measure the charge excess from the final-stage hadrons at the mid-rapidity. However, particle detectors usually have finite acceptance and tracking efficiency in momentum space and extrapolations of those particles to low momentum are different depending on the mass and collective effects expected to be present in heavy ion collisions. In addition, different interaction cross sections between the positive and negative hadrons of interest  with detector material complicates net-charge yield measurements. The situation is made worse due to the isospin balance in the finale state. For example, in $p+p$ collisions, one would expect that $\pi^-/\pi^+<1$ and $\bar{p}/p<1$ simply due to net positive charge of colliding protons. However, in A+A collisions~\cite{STAR:2008med}, one would expect $\pi^-/\pi^+>1$ and $\bar{p}/p<1$ due to the detailed balance of isospin from neutron excess and most of the stable colliding nuclei going through processes such as $n+X\leftrightarrow p+\pi^- + X$. The charge excesses in pions and baryons have opposite sign in A+A collisions and they would partly cancel. All of these complications have prevented previous experiments from obtaining a precise measurement of absolute charge stopping at mid-rapidity~\cite{Wong:2000cp, Stankus:2006zn}.

In 2018, RHIC delivered a set of isobar collisions of ${}^{96}_{44}$Ru+$^{96}_{44}$Ru and ${}^{96}_{40}$Zr+$^{96}_{40}$Zr at the center-of-mass energy per nucleon-nucleon pair of 200 GeV~\cite{STARCME:2021mii}. The data set collected by the STAR collaboration is of high statistics (2 Billion events/species) and quality. The isobar run was conducted in such a way that several of the aforementioned systematic uncertainties will be cancelled in the ratio of observables between the two isobar species. We propose that this can be used to study if valence quarks (reflected in the charge) and baryons are shifted to mid-rapidity from the beam rapidity (or stopped) in the same way. If the total charge and baryon number are shifted differently from beam rapidity to mid-rapidity, it would indicate that baryon number is not correlated with the valence quarks and may likely be carried by the baryon junction~\cite{XZB2022}. 

Measurements of net baryons at mid-rapidity can be performed using net protons as we discussed earlier. However, the absolute measurement of net charge is highly nontrivial. The net charge here is defined based on the yields of pion, kaon and (anti)proton: 
\begin{equation}
Q=(N_{\pi^+}+N_{K^+}+N_p)-(N_{\pi^-}+N_{K^-}+N_{\bar{p}}).
\end{equation}
We propose a method to precisely measure the net-charge difference at mid-rapidity between the two collision systems, defined as:
\begin{equation}
\Delta Q = Q^{\rm Ru} - Q^{\rm Zr},
\end{equation}
using double ratios. Specifically, the double ratio of pions can be defined as
\begin{equation}
\begin{split}
R2_{\pi} &= (N^{\rm Ru}_{\pi^+}/N^{\rm Ru}_{\pi^-})/(N^{\rm Zr}_{\pi^+}/N^{\rm Zr}_{\pi^-})\nonumber \\
 &= 1+ [(N^{\rm Ru}_{\pi^+}/N^{\rm Ru}_{\pi^-}) - (N^{\rm Zr}_{\pi^+}/N^{\rm Zr}_{\pi^-})]/(N^{\rm Zr}_{\pi^+}/N^{\rm Zr}_{\pi^-}), \nonumber \\
    &\simeq 1+ (N^{\rm Ru}_{\pi^+}/N^{\rm Ru}_{\pi^-}) - (N^{\rm Zr}_{\pi^+}/N^{\rm Zr}_{\pi^-}), 
\end{split}
\end{equation}
expanding to first order in $(N_{\pi^+}/N_{\pi^-}-1)$ in each collision system. Consequently, the pion contribution to $\Delta Q$ can be written as: 
\begin{equation}
\begin{split}
\Delta Q_{\pi}  &= Q_{\pi}^{\rm Ru} - Q_{\pi}^{\rm Zr}\\
&= (N^{\rm Ru}_{\pi^+}-N^{\rm Ru}_{\pi^-})-(N^{\rm Zr}_{\pi^+}-N^{\rm Zr}_{\pi^-})\\
&= N_{\pi}(R2_{\pi}-1).
\label{eq:delta_Q_pi}
\end{split}
\end{equation}
Here, the expansion approximation is again up to first order in $(N_{\pi^+}/N_{\pi^-}-1)$ assuming $\Delta Q_{\pi}<<Q_{\pi}^{\rm Ru}\sim Q_{\pi}^{\rm Zr}<<N_{\pi}$. The notation $N_{\pi}$ refers to the average yield between $\pi^+$ and $\pi^-$ and between Ru+Ru and Zr+Zr collisions: $N_{\pi}=(N_{\pi^+}^{\rm Ru} + N_{\pi^-}^{\rm Ru}+N_{\pi^+}^{\rm Zr} + N_{\pi^-}^{\rm Zr})/4$. Consequently, the net-charge difference can be expressed as: 
\begin{equation}
\begin{split}
\Delta Q &=\Delta Q_{\pi}+\Delta Q_{K}+\Delta Q_{p}\\
&= N_{\pi}(R2_{\pi}-1)+{N_K}(R2_K-1)+{N_p}(R2_p-1), 
\label{eq:delta_Q}
\end{split}
\end{equation}
where $R2_{K}$, $R2_{p}$, $N_K$ are $N_p$ are defined similarly as those for pions.

The derivation above assumes that the two isobar colliding systems would have had the same multiplicity at a given matching centrality. 
In reality, the multiplicity and centrality between two isobar systems do not match exactly~\cite{STARCME:2021mii}. 
In the following, we derive the vigorous expansion terms under the condition of small multiplicity mismatch between the two isobar systems and the approximation for uncertainty estimation. Let's define: 
$$N_{\pi^+}^{\rm Ru}=N_{\pi}^{\rm Ru}+\delta_1$$
$$N_{\pi^-}^{\rm Ru}=N_{\pi}^{\rm Ru}-\delta_1$$
$$N_{\pi^+}^{\rm Zr}=N_{\pi}^{\rm Zr}+\delta_2$$
$$N_{\pi^-}^{\rm Zr}=N_{\pi}^{\rm Zr}-\delta_2$$
The multiplicity difference between Ru+Ru and Zr+Zr collisions can be written in terms of small excess $\delta$: 
$$N_{\pi}^{\rm Ru}=N_{\pi}+\delta$$
$$N_{\pi}^{\rm Zr}=N_{\pi}-\delta$$
Effectively, we redefine the four pion measurements into four other variables ($N_{\pi}$, $\delta$, $\delta_1$ and $\delta_2$). Therefore, 
$$N_{\pi^+}^{\rm Ru}=N_{\pi}+\delta+\delta_1$$
$$N_{\pi^-}^{\rm Ru}=N_{\pi}+\delta-\delta_1$$ 
$$N_{\pi^+}^{\rm Zr}=N_{\pi}-\delta+\delta_2$$
$$N_{\pi^-}^{\rm Zr}=N_{\pi}-\delta-\delta_2$$
Since the multiplicities of the two isobar systems are different, it is incorrect to directly calculate the charge difference as the following: 
\begin{equation*}
\begin{split}
\Delta Q_{\pi}&=(N_{\pi^+}^{\rm Ru}
-N_{\pi^-}^{\rm Ru})-(N_{\pi^+}^{\rm Zr}-N_{\pi^-}^{\rm Zr})\\
&=2(\delta_1-\delta_2)
\end{split}
\end{equation*}
Alternatively, the charge difference should be defined as: 
\begin{equation*}
\begin{split}
\Delta Q_{\pi}&=(N_{\pi^+}^{\rm Ru}
-N_{\pi^-}^{\rm Ru}){\frac{N_{\pi}}{N_{\pi}+\delta}}-(N_{\pi^+}^{\rm Zr}-N_{\pi^-}^{\rm Zr}){\frac{N_{\pi}}{N_{\pi}-\delta}}\\
&={\frac{2N_{\pi}}{N_{\pi}^2-\delta^2}}(N_{\pi}(\delta_1-\delta_2)-\delta(\delta_1+\delta_2))\\
&\simeq2(\delta_1-\delta_2)-{\frac{2\delta}{N_{\pi}}}(\delta_1+\delta_2)-2\left(\frac{\delta}{N_{\pi}}\right)^3(\delta_1+\delta_2)+[...]
\end{split}
\end{equation*}
The first term in the last equation is the difference $(\delta_1-\delta_2)$ while the second term is the product of the sum $(\delta_1+\delta_2)$ and $\delta/N_{\pi}$. These two terms are of the same order of magnitude if the multiplicity mismatch ($\delta/N_{\pi}$) between the two isobar systems is on the order of a few percent or higher. 

On the other hand, the double ratio can be rewritten as:
\begin{equation*}
\begin{split}
R2_{\pi}&={\frac{(N_{\pi^+}^{\rm Ru}/N_{\pi^-}^{\rm Ru})}{(N_{\pi^+}^{\rm Zr}/N_{\pi^-}^{\rm Zr})}}\\
&={\frac{(N_{\pi^+}^{\rm Ru}\times N_{\pi^-}^{\rm Zr})}{(N_{\pi^+}^{\rm Zr}\times N_{\pi^-}^{\rm Ru})}}\\
&={\frac{(N_{\pi}+\delta+\delta_1)(N_{\pi}-\delta-\delta_2)}{(N_{\pi}-\delta+\delta_2)(N_{\pi}+\delta-\delta_1)}}\\
&={\frac{N_{\pi}^2+N_{\pi}(\delta_1-\delta_2)-(\delta+\delta_1)(\delta+\delta_2)}{N_{\pi}^2-N_{\pi}(\delta_1-\delta_2)-(\delta-\delta_1)(\delta-\delta_2)}}
\end{split}
\end{equation*}
Under the assumption of $\delta<< N_{\pi}$, $\delta_1<<N_{\pi}$ and $\delta_2<<N_{\pi}$, we can omit any higher-order terms of $(\delta_{,1,2}/N_{\pi})^3$ (order of $10^{-6}$) and obtain: 
\begin{eqnarray*}
R2_{\pi}\simeq 1+{\frac{2}{N_{\pi}}}(\delta_1-\delta_2)-{\frac{2\delta}{N_{\pi}^2}}(\delta_1+\delta_2)\\
+{\frac{2}{N_{\pi}^2}}(\delta_1-\delta_2)^2+\left(\frac{1}{N_{\pi}}\right)^3[...]+[...]
\end{eqnarray*}

One can see from the above equations that both $R2$ and  $\Delta Q$ are sensitive to the multiplicity difference between the two isobar collision systems ($\delta$) at the second order ($\delta(\delta_1+\delta_2$)). More importantly, the second and third terms in $R2_{\pi}$ coincide with the first and second terms of $\Delta Q_{\pi}$, and the relationship between $R2$ and $\Delta Q$ reduces to:
$$R2_{\pi}=1+\Delta Q_{\pi}/N_{\pi}$$
This approximation ignores higher order contribution at the level of $<1\%$ of $\Delta Q_{\pi}$. Finally: 
$$\Delta Q_{\pi}=N_{\pi}(R2_{\pi}-1)$$
, which is the same as Eq. \ref{eq:delta_Q_pi}. This is to say that Eq. \ref{eq:delta_Q}, based on particle yields and double ratios, can still be used to calculate net-charge difference between the two isobar collision systems even if there is multiplicity mismatch.

\begin{figure}[htb]
  \begin{center}
    \includegraphics[width=0.49\textwidth]{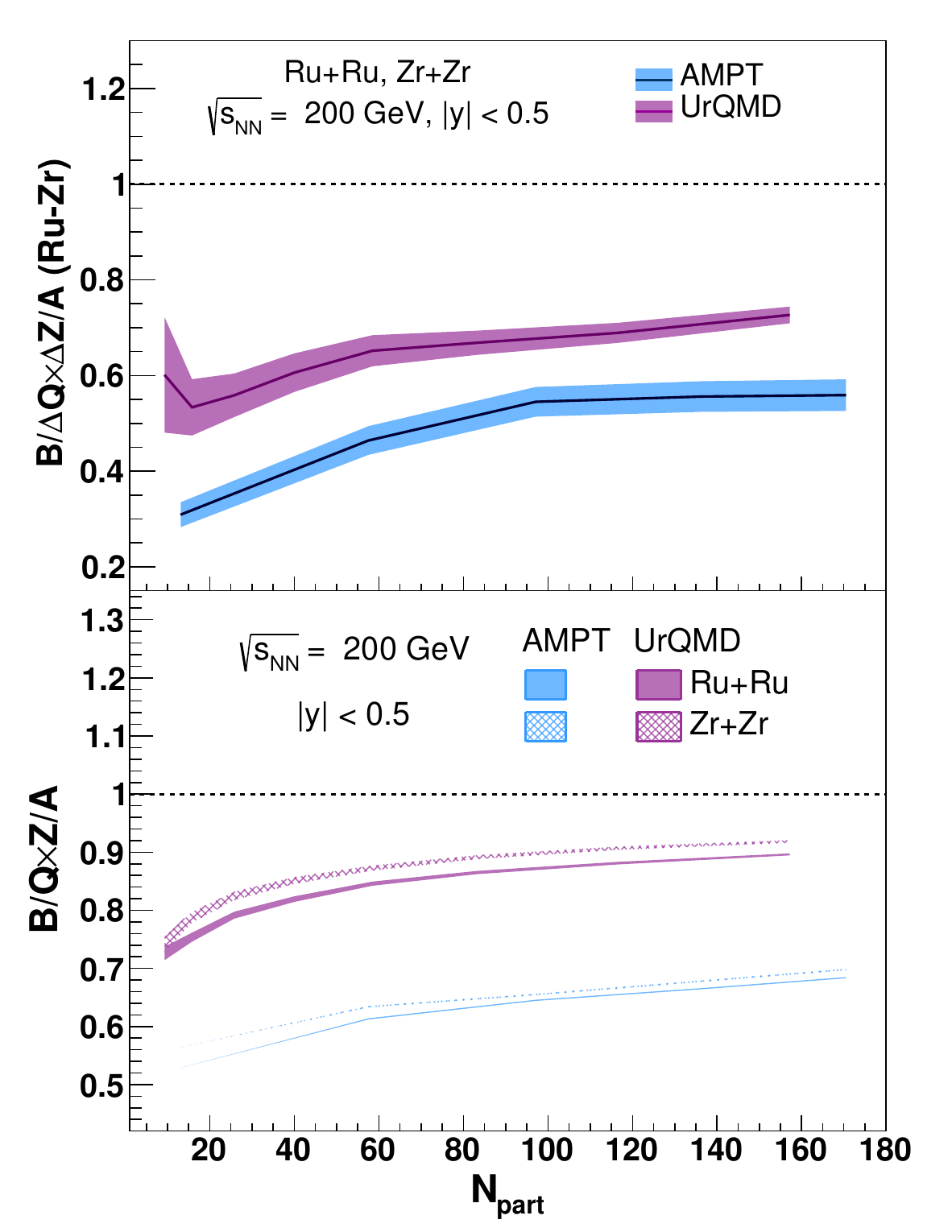}
  \end{center}
  \caption{\label{fig_BtodQ_model} Distributions of $B/\Delta Q\times\Delta Z/A$ (top) and $B/Q\times Z/A$ (bottom) as a function of $N_{\rm{part}}$ in 200 GeV Ru+Ru and Zr+Zr collisions predicted by AMPT and UrQMD models~\cite{bleicher1999relativistic,Lin:2004en}.}
\end{figure}

How does $\Delta Q$ compare to the expected charge difference if all the baryon stopping also shifts the valence quarks? The top panel of Fig. \ref{fig_BtodQ_model} shows the distributions of $B/\Delta Q\times\Delta Z/A$ at mid-rapidity ($|y| < 0.5$) between Ru+Ru and Zr+Zr collisions as a function of centrality, indicated by the number of participating nucleons ($N_{\rm{part}}$), from AMPT and UrQMD model calculations~\cite{bleicher1999relativistic,Lin:2004en}. Here $\Delta Z = 44 - 40 = 4$, and $A = 96$, common to the two isobars. In both AMPT and UrQMD models, the baryon number is carried by valence quarks. The quantity $\Delta Q^{\prime}=B\times\Delta Z/A$ represents the expected charge stopping difference between Ru+Ru and Zr+Zr collisions if net-baryon and net-charge stoppings were exactly the same. As evident from model calculations, $\Delta Q^{\prime}/\Delta Q$ at mid-rapidity is less than one in all centrality classes. Results from the string melting AMPT model show that the deviation from the naive expectation of one is mostly due to the asymmetry in the strange quark rapidity distributions, {\it i.e.}, there are more anti-strange quarks at mid-rapidity than strange quarks, resulting in more charge stopping than baryon stopping. This could be due to that strange quarks are pulled away along with valence quarks at large rapidity since it is easier for them to form baryons when close together. These calculations provide a baseline for experimental search of the baryon junction by comparing charge and baryon stopping. One of the remaining questions is whether the absolute charge transport to mid-rapidity is directly proportional to the charge difference between two isobar ions. Although the answer may depend on the exact baryon transport mechanism we seek to address and is to date unkown, we could use UrQMD and AMPT models to extract the absolute charge transports for the two isobar collisions separately. The bottom panel of Fig.~\ref{fig_BtodQ_model} shows $B/Q\times Z/A$ from both models and the results are qualitatively consistent with $B/\Delta Q\times \Delta Z/A$ between two isobar ions shown in the top panel.

\section{Comparison with event generators}
In this section, we compare calculations from various event generators to measurements of proton and antiproton production at mid-rapidity in $p+p$ collisions of different energies. Specifically, the following event generators and tunes are tested: PYTHIA 6.4 (default tune, Perugia0 (P0) tune, Perugia2012 (P12) tune) \cite{Sjostrand:2006za, Skands:2010ak}, PYTHIA 8.1 (default tune) \cite{Sjostrand:2007gs}, PYTHIA 8.3 (CR Mode 2) \cite{Bierlich:2022pfr, Christiansen:2015yqa} and HERWIG 7.2 \cite{Bahr:2008pv, Bellm:2015jjp}. For PYTHIA 6.4, the hard QCD processes are simulated, while for PYTHIA 8.1 and 8.3, non-diffractive events are used. On the other hand, the HERWIG simulation uses QCD $2 \rightarrow 2$ processes with the matrix element ``MEQCD2to2". Both PYTHIA 6.4 and PYTHIA 8.1 default tune produce baryons mainly through the ``popcorn" mechanism while the PYTHIA 8.3 CR tune \cite{Christiansen:2015yqa} allows additional string junctions to be created through color reconnections to enhance the baryon production at mid-rapidity. Meanwhile, HERWIG uses cluster hadronization model to form hadrons. For all simulations, resonance and weak decays are turned on. 
 
\begin{table*}[tbh]
\begin{tabularx}{0.99\textwidth} { 
  | >{\centering\arraybackslash}X 
  | >{\centering\arraybackslash}X 
  | >{\centering\arraybackslash}X
  | >{\centering\arraybackslash}X
  | >{\centering\arraybackslash}X
  | >{\centering\arraybackslash}X| }
 \hline
  & $\bar{p}/p$ (200 GeV, $|y| < 0.1$) & $\bar{p}/p$ (900 GeV, $|y| < 0.5$) & $\bar{p}/p$ (7 TeV, $|y| < 0.5$) & $dN_{p-\bar{p}}/dy$ slope & $B/Q$ (200 GeV, $|y| < 0.5$)  \\
 \hline
 Data  & $0.819\pm0.047$ & ($0.957\pm0.015$) & ($0.991\pm0.015$) & $0.61\pm0.03$ & -  \\
 \hline
PYTHIA 6.4  & $0.928\pm0.008$ & $0.966\pm0.003$ ($0.961\pm0.004$) & $0.986\pm0.003$ ($0.984\pm0.003$) & $0.86\pm0.05$  & $0.300\pm0.016$   \\
PYTHIA 6.4 (P0) & $0.851\pm0.008$ & $0.948\pm0.004$ ($0.946\pm0.004$) & $0.987\pm0.003$ ($0.985\pm0.004$) & $0.76\pm0.03$ & $0.879\pm0.049$ \\
PYTHIA 6.4 (P12) & $0.815\pm0.007$ & $0.912\pm0.003$ ($0.911\pm0.004$) & $0.966\pm0.003$ ($0.968\pm0.003$) & $0.38\pm0.02$ & $1.519\pm0.095$  \\
PYTHIA 8.1 & $0.913\pm0.008$ & $0.968\pm0.003$ ($0.964\pm0.004$) & $0.989\pm0.003$ ($0.987\pm0.003$) & $0.89\pm0.04$ & $0.567\pm0.032$ \\
PYTHIA 8.3 (CR Mode 2) & $0.837\pm0.006$ & $0.928\pm0.003$ ($0.926\pm0.003$) & $0.981\pm0.002$ ($0.982\pm0.003$) & $0.73\pm0.02$ & $0.986\pm0.033$ \\
HERWIG 7.2 & $0.858\pm0.005$ & $0.969\pm0.002$ ($0.914\pm0.003$) & $1.028\pm0.002$ ($1.006\pm0.002$) & $1.17\pm0.04$ & $0.558\pm0.014$ \\
\hline
\end{tabularx}
  \caption{\label{tab_EventGenerator} Comparisons between data \cite{STAR:2008med, ALICE:2010hjm} and various event generators and tunes \cite{Sjostrand:2006za, Skands:2010ak, Sjostrand:2007gs, Bierlich:2022pfr, Christiansen:2015yqa, Bahr:2008pv, Bellm:2015jjp} for quantities related to baryon stopping, as well as model predictions for $B/Q$ in 200 GeV $p+p$ collisions. Values in parentheses are for primordial production, while others are for inclusive production. Listed uncertainties are the quadrature sums of statistical and systematic uncertainties for data, and statistical only for simulations.}
\end{table*}

Measurements of antiproton to proton ratio at mid-rapidity in $p+p$ collisions at $\sqrt{s}$ = 200 GeV \cite{STAR:2008med}, 900 GeV and 7 TeV \cite{ALICE:2010hjm} are listed in Tab. \ref{tab_EventGenerator}. The measurement at 200 GeV is for inclusive production including weak decays, while the other two are for primordial production. Model calculations for both inclusive and primordial, shown in parentheses, production are listed for comparison. For PYTHIA simulations, the $\bar{p}/p$ ratios are compatible within statistical errors between inclusive and primordial production, while for HERWIG, the primordial ratio is significantly lower than that for the inclusive production. Simulations are generally in agreement with data except that PYTHIA 6.4 P12 tune and HERWIG 7.2 underestimate the primordial $\bar{p}/p$ ratio at 900 GeV by about 3$\sigma$. The rapidity dependence of inclusive net-proton yield is obtained similarly as in data (Figs. \ref{fig_baryon} and \ref{fig_baryon_allcent}), {\it i.e.}, simulating $p+p$ collisions from low to high energies and calculating the net-proton yields at mid-rapidity. The yield dependence on beam rapidity is fit with exponential functions in the energy range of 30 - 200 GeV for PYTHIA and 80 - 200 GeV for HERWIG, and the resulting slopes are shown in Tab. \ref{tab_EventGenerator}. They are compared to that extracted from heavy-ion collisions as presented in Fig. \ref{fig_baryon}. As shown in Fig. \ref{fig_baryon_allcent}, the slope parameter does not depend on collision centrality, and therefore one can make such a comparison between $p+p$ and heavy-ion collisions. The PYTHIA 6.4 P12 tune significantly underestimates the slope, while HERWIG overshoots data by almost a factor of 2. Finally, predictions on the ratio of charge and baryon stopping at mid-rapidity ($B/Q$) in $p+p$ collisions at $\sqrt{s}$ = 200 GeV from different event generators and tunes are also presented. It is worth noting that the $B/Q$ ratio spans a very large range between different predictions. Future precise measurement of $B/Q$, in combination with existing results on $\bar{p}/p$ and $dN_{p-\bar{p}/}/dy$ slope, will provide stringent tests on the baryon production mechanism implemented in models. 

\section{Opportunities with future RHIC runs and EIC}

Recently collected data and remaining years of RHIC runs along with the extended pseudorapidity reach offered by the recent upgrade of the STAR detector will provide unique opportunities for future measurements with ion energies of $E_A$=100 GeV per nucleon. The forward and the mid-rapidity upgrade program of STAR that includes: 1) the inner Time Projection Chamber (iTPC, $-1.5<\eta<1.5$)~\cite{itpc}, 2) highly granular forward Event-Plane Detectors (EPD, $2.1<|\eta|<5.1$)~\cite{Adams:2019fpo} and, 3) newly installed forward tracking and calorimetry system (FTS $\&$ FCS, $2.5<\eta<4$)~\cite{STARnote:648}. With the combination of these three sub-systems the asymmetric pseudorapidity distributions of charged hadrons in $\gamma$+Au interactions can be captured over six units of pseudorapidity. This will improve the trigger purity and provide a wider range of rapidity for net-baryon measurements. An anticipated $p$+Au run of RHIC in the year 2024 will provide an opportunity to collect a high statistics sample of $\gamma+p$ processes. The analysis of such a data set can provide data-driven baseline for measurements in $\gamma+$Au processes at the same collision energy.

In our study, we use event generators in e+A and e+p collisions to study the photonuclear process. It is clear that such semi-inclusive processes could be cleanly identified and analyzed in the future EIC since there are no background nucleus-nucleus collisions. The primary detector at the EIC, ePIC, will include barrel and endcap Time-of-Flight detectors that have low-momentum capability for baryon measurements over a wide range of rapidity in e+p and e+A collisions; therefore, they are ideal for performing the rapidity dependence of baryons~\cite{AbdulKhalek:2021gbh}. Moreover, with the possibility of a second detector at the EIC, which has better stable acceptance of low $Q^2$, exciting opportunities are possible to study baryon transport~\cite{det2xu}. We propose to perform e+Ru and e+Zr collisions at the EIC, and measure the net-charge and net-baryon as a function of $x$ and $Q^2$ and test the baryon junction picture in an approach similar to what is proposed for isobar collisions at RHIC~\cite{det2niseem}. In a follow-up paper, we plan to explore this in detail. In addition, the dependence of the stopping with photon virtuality $\gamma^*+A\rightarrow X+p$ can be performed in the future EIC. However, it does require that the detectors can cleanly identify protons and antiprotons at low momentum. Our proposal will also complement measurements of backward photoproduction of mesons at the EIC in exclusive processes such as $\gamma^*+p\rightarrow \omega+p$ at far-forward rapidity~\cite{spencer2019,Gayoso:2021rzj,PhysRevLett.123.182501}. For a detailed discussion on the connection of such processes with baryon stopping, we refer the readers to Ref~\cite{Cebra:2022avc}. 

\section{Summary}
Baryon number is one of the best known and stringently tested conserved quantities in physics, and it could be carried by a gluon topology instead of quarks.  Many experimental results cannot be explained by the conventional models and suggest that gluon junctions may play a significant role in the baryon stopping experimentally observed in rapidity distributions in central A+A, isobar, and $\gamma+$A collisions. We presented three possible observables that may shed light into what carries this quantum number: is it quarks or a gluonic topological junction? Future data analyses and experiments with the proposed observables would provide conclusive answers to this fundamental question. 

\section{acknowledgements} 
The authors thank Zhouduming Tu for assistance with the BeAGLE event generator, Mriganka M. Mondal and Kolja Kauder for help with PYTHIA simulation, Fuqiang Wang for an earlier version of Fig.~\ref{fig_baryon}, the STAR Collaboration for information about detector capability, and, Dmitri Kharzeev,  Spencer Klein, Volker Koch, Wenliang Li, Krishna Rajagopal, Bj\"orn Schenke, Chun Shen, Vladimir Skokov, Mark Strikman, Zachary Sweger for fruitful discussions. Brandenburg's work was supported in part by the Department of Energy early Career award. This work was funded by the U.S. DOE Office of Science under contract No. DE-sc0012704, DE-FG02-10ER41666, DE-AC02-98CH10886, and the National Science Foundation 
under Grant No. 2012947 and 2310021. 
\bibliography{bibliography}

\end{document}